\documentclass[%
 reprint,
superscriptaddress,
 amsmath,amssymb,
 aps,
 prl,
]{revtex4-2}

\usepackage{graphicx}
\usepackage{dcolumn}
\usepackage{bm}
\usepackage{mathptmx}
\usepackage{siunitx}
\usepackage{amssymb}
\DeclareMathAlphabet{\mathcal}{OMS}{cmsy}{m}{n}

\begin{document}


\title{Universal Particle Kinetic Distribution in Crowded Environments}

\author{Dian Fan}
\email{d.fan@ucl.ac.uk}
\affiliation{Department of Chemical Engineering, University College London, London WC1E 7JE, UK.}

\author{Ronny Pini}
\affiliation{Department of Chemical Engineering, Imperial College London, London SW7 2AZ, UK.}

\author{Alberto Striolo}
\email{a.striolo@ucl.ac.uk}
\affiliation{Department of Chemical Engineering, University College London, London WC1E 7JE, UK.}

\date{\today}

\begin{abstract}
We study many-particle transport in heterogeneous, crowded environments at different particle P\'{e}clet numbers ($Pe^*$). We demonstrate that a modified Nakagami-$m$ function describes particle velocity probability distributions when particle deposition occurs. We assess the universality of said function through comparison against new Lagrangian simulations of various particle types as well as experimental data from the literature. We construe the function's physical meaning as its ability to explain particle deposition in terms of $Pe^*$ and the competition between distributions of energy barriers for particle release and particles' diffusive energy. 
\end{abstract}

\maketitle

\paragraph{Introduction.}{\hspace{-1em}}---Many-particle systems in crowded environments (packed with many obstacles) are ubiquitous in Nature, yielding a wealth of important phenomena ranging from clustering in granular gases \cite{RN1}, localization transition during colloidal gelation \cite{RN2}, biofilms formation \cite{RN3}, solute dispersion \cite{RN4} as well as nanoparticles deposition in porous media \cite{RN5,RN6}. Understanding the particles' kinetic probability distribution in real environments is essential to achieve particulate control \cite{RN7}. Prior studies described how such distributions depend on environmental variables (e.g., obstacle size \cite{RN8}, porosity \cite{RN9,RN10,RN11}, pore structure \cite{RN12}) by fitting the probability data into standard probability distribution functions (PDFs), such as exponential \cite{RN8,RN13,RN14}, stretched exponential \cite{RN9,RN15}, power-law \cite{RN12,RN16}, and power-exponential ones \cite{RN10}. However, the assessment of their predictive ability remains elusive, because these functions contain fitting parameters that often lack a solid physical foundation.

Phenomenologically, laboratory particle-tracking data for solutes, nanoparticles (NPs), and microparticles (MPs) through random packs of spheres manifest similar non-Gaussian velocity probability distributions \cite{RN3,RN8,RN17}, suggesting that a universal function might be able to describe particle kinetics. Should such universal function exist, it should be consistent with rigorous mathematical derivations, such as the Maxwell-Boltzmann (MB) distribution for non-interacting ideal gas particles \cite{RN18}. However, when particles transport through real environments, they can deposit \cite{RN19}, in which case the MB distribution is likely to break down \cite{RN20}. This observation calls for new theoretical developments to describe statistical particles' kinetics in crowded environments.

In this Letter, we consider many particles in a pressure-gradient-induced fluid flow through an obstacle-packed heterogeneous environment. We derive a universal function able to describe the particles' velocity distribution by modeling particles' behaviors near the pore center and the pore wall, respectively. The theory proposed rationalizes that particle P\'{e}clet number and the strength of particle-wall interactions govern particle kinetic distributions.

\paragraph{Particles near the pore center.}{\hspace{-1em}}---Let us consider the longitudinal transport of particles driven by a pressure gradient $(\nabla P)$ in a porous medium where the average pore size is much greater than the particle size. Particles near the pore center are thus unaffected by particle-wall interactions. In these conditions, the particle longitudinal velocity ($v_{\mathrm{pL}}$) can be decomposed into advective ($u_{\mathrm{pL}}$) and diffusive components ($w_{\mathrm{pL}}$), i.e.,  $v_{\mathrm{pL}}=u_{\mathrm{pL}}+w_{\mathrm{pL}}$. The (longitudinal) total energy of each particle ($E=m_{\mathrm{p}} v_{\mathrm{pL}}^{2}/2$) can therefore be expressed as a sum of advective ($K_{\mathrm{A}}=m_{\mathrm{p}}u_{\mathrm{pL}}^{2}/2$) and diffusive kinetic energy ($K_{\mathrm{D}}=m_{\mathrm{p}} w_{\mathrm{pL}}^{2}/2$), i.e., $E=K_{\mathrm{A}}+K_{\mathrm{D}}$ \cite{RN21}. Of note, we only consider positive, longitudinal velocities and their energies in this study.

Due to the imposed $\nabla P$, particles at location $x$ experience a flow potential $\varphi(x)=-\nabla Px/\bar{n}m_{\mathrm{p}}$ \cite{RN22}, where $\bar{n}$ is the mean particle number density, and $m_{\mathrm{p}}$ is the mean particle mass. At time $t \rightarrow+\infty$, the local particle number density, $n(x)$, obeys a Boltzmann distribution: $n(x)=n(x=0) \times\exp\left(\nabla Px/\bar{n}k_{\mathrm{B}}T\right)$, where $k_{\mathrm{B}}$ is the Boltzmann constant and $T$ is the absolute temperature \cite{Schekochihin}. By coupling $n(x)$ with the local MB distribution \cite{RN23}, we derive (see Supplemental Material (SM), Sec. II.A) the velocity PDF for particles near the pore center:
\begin{equation} \label{eq:1}
    f_{V}\left(v_{\mathrm{pL}}\right) \propto \exp \left(-\frac{-\frac{\nabla P}{\bar{n}} x+\frac{1}{2} m_{\mathrm{p}} w_{\mathrm{pL}}^{2}}{k_{\mathrm{B}} T_{1}}\right) \propto \exp \left(-\frac{m_{\mathrm{p}} v_{\mathrm{pL}}^{2}}{2 k_{\mathrm{B}} T_{1}}\right),
\end{equation}
where $T_{1}$ is the temperature that reflects the mean $\langle E\rangle=m_{\mathrm{p}}\langle v_{\mathrm{pL}}^{2}\rangle /2=k_{B}T_{1}/2$. Equation \eqref{eq:1} considers that after the particles are introduced at the upstream of the porous medium (at $x=0$) with zero velocity, their flow potential decreases as the advective kinetic energy increases, i.e., $\quad-\nabla Px/ \bar{n} = m_{\mathrm{p}} u_{\mathrm{pL}}^{2}/2$. 

\paragraph{Particles near the pore wall.}{\hspace{-1em}}---For particles near the pore wall (i.e., obstacle surfaces), it is assumed that advective velocity is negligible (i.e., $u_{\mathrm{pL}}=0$) because of fluid stagnation. The particles' longitudinal velocity near the pore wall ($v_{\mathrm{pL}}$) is the sum of the diffusive velocity ($w_{\mathrm{pL}}$) and an additional velocity term ($\theta$) affected by the particle-wall interactions: $v_{\mathrm{pL}}=w_{\mathrm{pL}}+\theta$. To quantify such particle-wall interactions, we adopt the extended Derjaguin-Landau-Verwey-Overbeek (XDLVO) formalism, where the net interaction energy is due to the combination of London-van der Waals (vdW), electric double layer, and short-range Born interactions \cite{RN24}. A typical XDLVO curve of net interaction energy versus separation distance features a deep primary minimum and a shallow secondary minimum, separated by a repulsive maximum (see SM, Sec. II.B). The difference between the local interaction energy and the repulsive energy maximum is the energy barrier ($\Delta E$), which hinders deposited particles from leaving the primary minimum. Therefore, $\Delta E$ is regarded as an activation energy for particle release \cite{Datta}. Once escaped from the primary minimum, depending on local thermal conditions, particles may escape from the secondary minimum, becoming free particles that transport along with the fluid stream.

To quantify the impact of the activation energy on $v_{\mathrm{pL}},$ we adopt the Arrhenius Equation \cite{RN19}, i.e.,
\begin{equation} \label{eq:2}
    v_{\mathrm{pL}}=v_{0} \exp \left(-\frac{\Delta E}{k_{\mathrm{B}} T}\right),
\end{equation}
where $v_{0}$ is the escaping velocity when the energy barrier is absent, i.e., $\Delta E=0$. 

In a real environment, where the ion distribution on the pore wall and the particles' instantaneous positions are inhomogeneous, the probability distribution of $\Delta E$ matters \cite{RN25,RN26}. Considering a typical disordered medium obtained by randomly packed spheres, $\Delta E$ can be described by a Boltzmann distribution \cite{RN18}, i.e., $f_{E}(\Delta E)=(k_{\mathrm{B}}T_{0})^{-1}\exp \left(-\Delta E/k_{\mathrm{B}}T_{0}\right)$, where $T_{0}$ is the temperature that represents the mean $\langle\Delta E\rangle=\int_{0}^{+\infty} \Delta E f_{E}(\Delta E)\mathrm{d} \Delta E=k_{\mathrm{B}}T_{0}$.

Based on $f_{V}(v_{\mathrm{pL}})=f_{E}(\Delta E)\left|\mathrm{d} \Delta E / \mathrm{d} v_{\mathrm{pL}}\right|$ \cite{RN27}, we derive the velocity PDF of particles near the pore wall as $f_{V}(v_{\mathrm{pL}})=\lambda v_{\mathrm{pL}}^{\lambda-1}/v_{0}^{\lambda}$, where $0\leq v_{\mathrm{pL}}\leq v_0$ and $\lambda=T/T_{0}$. 

Considering that particles, if not pre-existent in situ, must have transported to the near-wall region, the velocity probability of particles near the wall is, therefore, a conditional probability given that event of particles reaching the near-wall region has occurred (whose probability is denoted by $\mathcal{P}$). Therefore, the conditional probability is $f_{V \mid \mathcal{P}}\left(v_{\mathrm{pL}}\right)=f_{V}(v_{\mathrm{pL}}) / \mathcal{P}$, where we model $\mathcal{P}$ by interception efficiency \cite{Weber,RN24} for non-pre-existent particles in Eq. \eqref{eq:3} (also see SM, Sec. II.B), and $\mathcal{P}=1$ for pre-existent particles near the wall:
\begin{equation} \label{eq:3}
\mathcal{P} \approx \left\{
\begin{aligned}
    &\left(\frac{d_{\mathrm{p}}}{l}\right)^{q}\left(\frac{\mathcal{D}_{\mathrm{w}} Pe^{*}}{\kappa_{\mathrm{w}}}\right)^{2-q} & (\text{non-pre-existent particles}) \\
    &1 &(\text{pre-existent particles})
\end{aligned}
\right.,
\end{equation}
where $q=1.57$, $d_{\mathrm{p}}$ is particle size, $l$ is average pore size, $\kappa_{\mathrm{w}}$ is kinematic viscosity of the ambient fluid, and $\mathcal{D}_{\mathrm{w}}$ is molecular diffusion coefficient of the ambient fluid. 

By imposing $\int_{0}^{v_{0}} f_{V \mid \mathcal{P}}(v_{\mathrm{pL}})\mathrm{d} v_{\mathrm{pL}}=1$, we obtain the conditional PDF of velocities near the pore wall:
\begin{equation} \label{eq:4}
    f_{V \mid \mathcal{P}}\left(v_{\mathrm{pL}}\right)=\frac{\alpha}{v_{0}^{\alpha}} v_{\mathrm{pL}}^{\alpha-1},
\end{equation}
where $\alpha=\lambda/\mathcal{P}=T/\mathcal{P}T_{0}$; the physical significance of $\alpha$ will be elaborated later.

\paragraph{The universal function.}{\hspace{-1em}}---Under the assumption that particles near the pore wall and near the pore center behave independently from each other, we derive the joint PDF for the longitudinal velocity of the entire particle ensemble via Eqs. \eqref{eq:1} and \eqref{eq:4}: $f_{V}(v_{\mathrm{pL}})=\beta v_{\mathrm{pL}}^{\alpha-1} \exp (-\gamma v_{\mathrm{pL}}^{2})$. Parameters $\beta$ and $\gamma$ are determined by imposing that  $\int_{0}^{+\infty} f_{V}(v_{\mathrm{pL}}) \mathrm{d} v_{\mathrm{pL}}=1$ and $\int_{0}^{+\infty} f_{V}(v_{\mathrm{pL}}) v_{\mathrm{pL}}^{2} \mathrm{d} v_{\mathrm{pL}}=\langle v_{\mathrm{pL}}^{2}\rangle$.

After defining a dimensionless velocity $v=v_{\mathrm{pL}} /\langle v_{\mathrm{w}}\rangle,$ where $\langle v_{\mathrm{w}}\rangle$ is the mean fluid velocity, we derive (see SM, Sec. II.C) the velocity PDF $(\forall v \geq 0)$ for the entire particle ensemble:
\begin{equation} \label{eq:5}
    f_{V}(v)=\frac{2 m^{m}}{\Gamma(m) \Omega^{m}} v^{2 m-1} \exp \left(-\frac{m}{\Omega} v^{2}\right),
\end{equation}
where $m=\alpha/2$ and $\Omega=\langle v^{2}\rangle$.

Equation \eqref{eq:5} manifests as a Nakagami-$m$ distribution, in which case the parameters $m$ and $\Omega$ are statistically referred to as shape and scale factors, respectively. The Nakagami-$m$ distribution was originally introduced to describe the fading signal intensity in wireless communications, which is characterized by $m \geq 0.5$ \cite{RN28}. Because we show later that in particulate systems $m$ can be $<0.5$, we refer to Eq. \eqref{eq:5} as a \textit{modified Nakagami-$m$ distribution} in the remainder of the Letter.

By letting $v \rightarrow+\infty$ and $v \rightarrow 0,$ we find that the PDF in Eq. \eqref{eq:5} for fast and slow particles is proportional to $\exp \left(-m v^{2} / \Omega\right)$ and $v^{2 m-1},$ respectively, which are proportional to Eqs. \eqref{eq:1} and \eqref{eq:4}, respectively, indicating that the correspondent particles are those near the pore center and the pore wall, respectively. The rest of the particles that are located between pore centers and walls, occupying a large portion of the entire ensemble, have intermediate velocities ($1<v<5$, see SM, Sec. VII) that represent transition states from immobile (i.e., slow) to highly mobile (i.e., fast). Based on our analysis, their velocity distributions are not asymptotic to $v^{2m-1}$ nor to $\exp \left(-m v^{2} / \Omega\right),$ but to the modified Nakagami-$m$ PDF (see SM, Sec. VII). Therefore, given its ability to capture the velocity distribution of the \textit{entire velocity range}, the modified Nakagami-$m$ distribution appears to be a unique, all-inclusive function for describing particle kinetics.

\paragraph{Numerical assessment of the universality of the modified Nakagami-$m$ distribution.}{\hspace{-1em}}---To assess both the uniqueness and the universality of the modified Nakagami-$m$ distribution (Eq. \eqref{eq:5}), we perform numerical simulations of particles transport through a three-dimensional randomly jammed packing of spheres (obstacles). The transporting particles considered include uranine (solute), nanoparticles (NPs), and microparticles (MPs). The relative importance of advection versus diffusion in establishing the particulate flow distribution and the fluid flow distribution are summarized by the particle P\'{e}clet number $Pe^{*}=\langle v_{\mathrm{w}}\rangle d_{\mathrm{p}} / \mathcal{D}_{\mathrm{w}}$ \cite{RN29}, and the pore P\'{e}clet number $Pe=\langle v_{\mathrm{w}}\rangle l / \mathcal{D}_{\mathrm{w}}$ \cite{Bijeljic}, respectively. In our numerical simulations, particles are subject to hydrodynamic drag, diffusive, and gravity (buoyancy) forces, particle-particle interactions (Coulomb and vdW forces), and particle-wall interactions. In each simulation, 3000 particles of the same type are studied. Computational details, including the validation of the algorithm, are presented in SM, Secs. IV and V \cite{Seguin,Skoge,Zhang,Morales,Ma,Mikutis,Happel,Holzner}.  
\begin{figure}[t]
\includegraphics[width=0.48\textwidth]{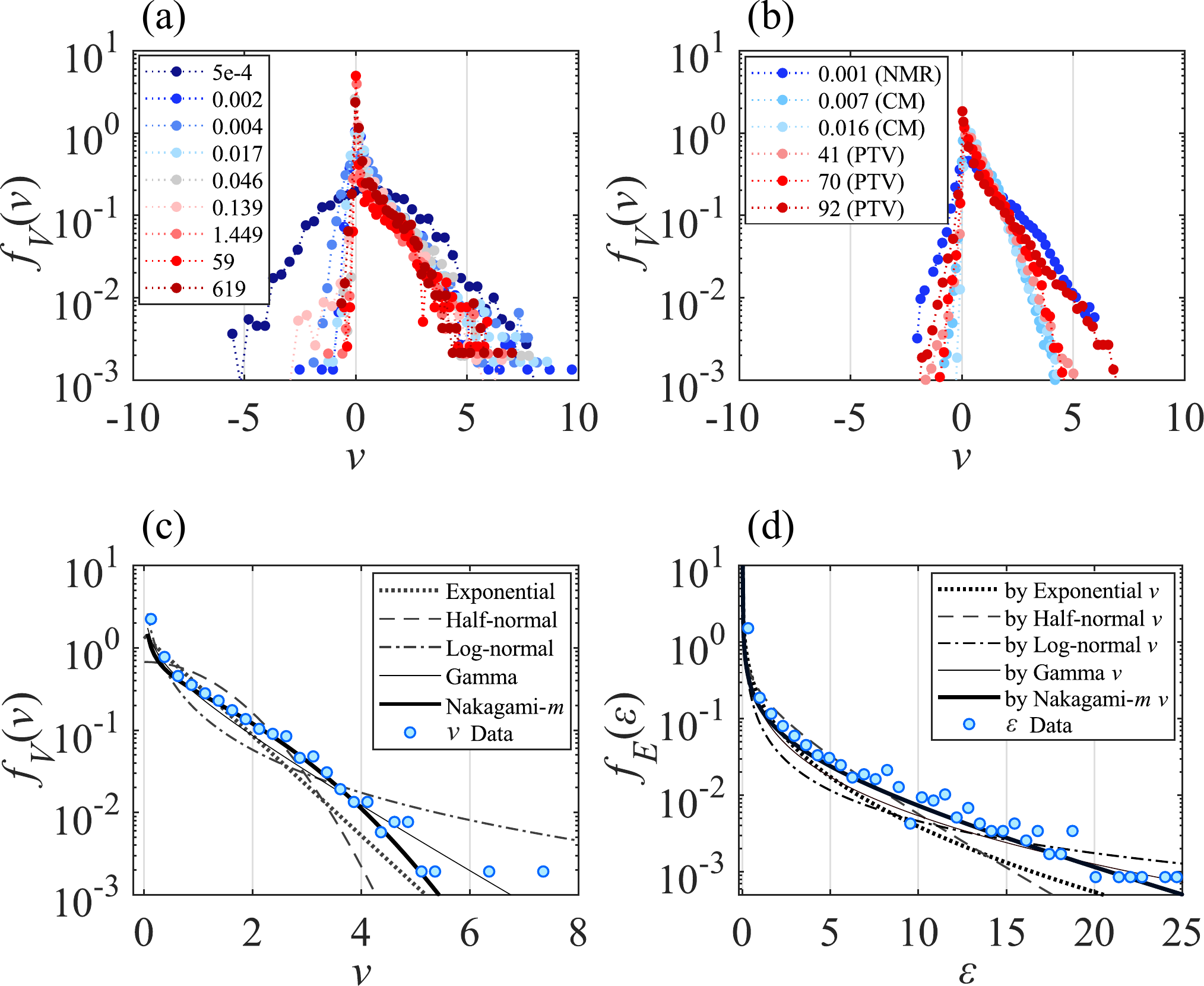}
\caption{\label{fig:1} (a) Probability distributions of $v$ obtained by our Lagrangian simulations at different $Pe^{*}$ (see legend). (b) Experimental probability distributions of $v$ measured at different $Pe^{*}$ via nuclear magnetic resonance (NMR) \cite{RN17}, confocal microscopy (CM) \cite{RN8}, and particle tracking velocimetry (PTV) \cite{RN3}. (c) ML estimations of the modified Nakagami-$m$ and other PDFs (see legend) based on the simulated $v$ data (symbols) at $Pe^{*}=59$ (i.e., MP at $Pe=1027$). (d) PDF predictions of $\varepsilon$ by the modified Nakagami-$m$ versus other $v$ PDFs (see legend) are compared to simulated $\varepsilon$ data (symbols) at $Pe^{*}=59$ (i.e., MP at $Pe=1027$).}
\end{figure}

\begin{figure}[t]
\includegraphics[width=0.45\textwidth]{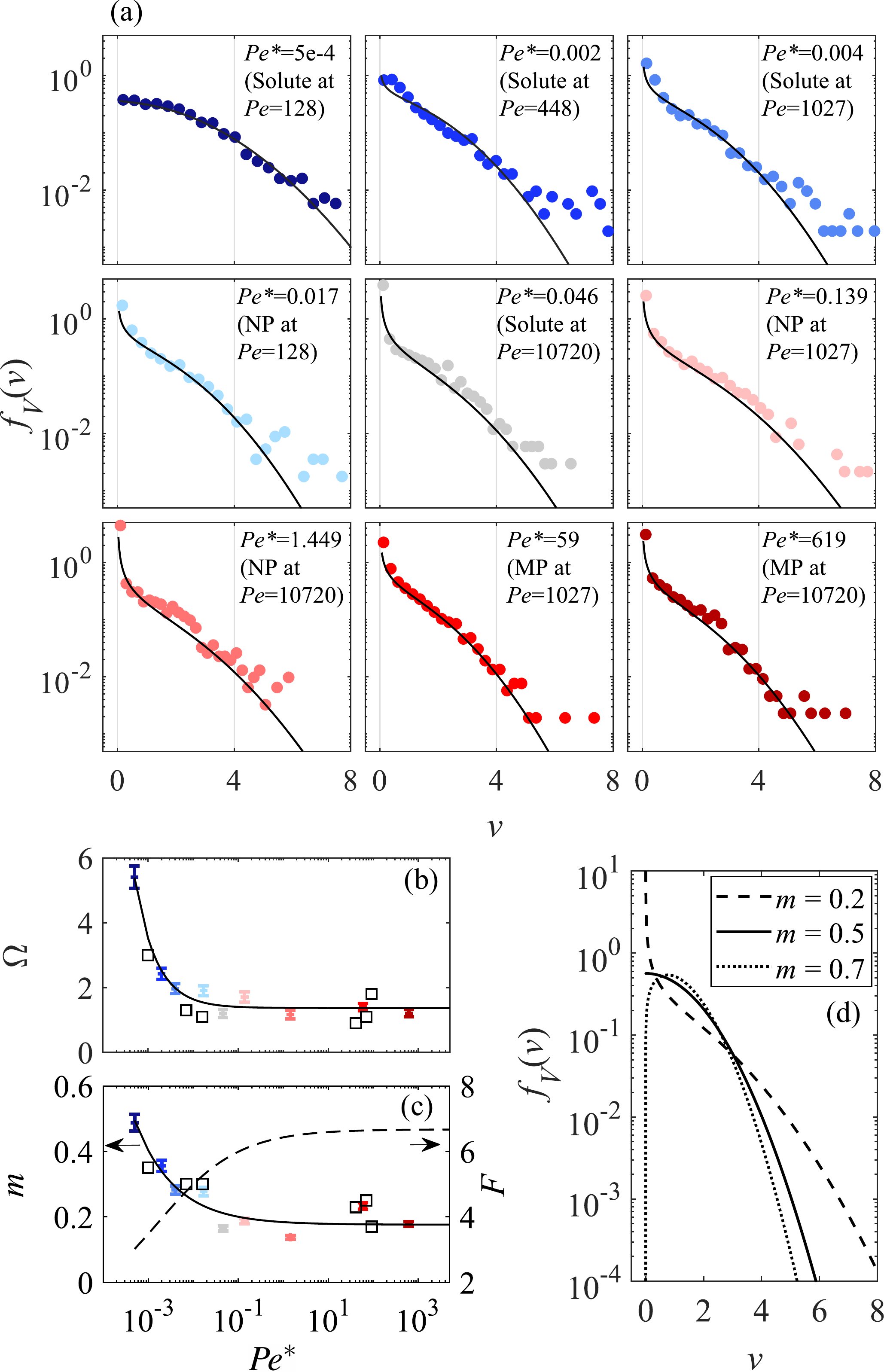}
\caption{\label{fig:2} (a) Modified Nakagami-$m$ fit (solid line) of $v$ data (symbols) from Lagrangian simulations at increasing $Pe^{*} \in\left[5\times10^{-4}, 619\right]$ (The simulation data are the positive $v$ in Fig. \ref{fig:1}(a)). Estimates of (b) $\Omega$ and (c) $m$ from simulations (symbols with 95\% confidence interval error bars) and experiments (squares) from Fig. \ref{fig:1}(b). The power-law scalings (solid lines) are obtained by fitting simulation data as a function of $Pe^{*}:\Omega=0.004239 \cdot\left(Pe^{*}\right)^{-0.9019}+1.37$ and $m=0.007877 \cdot\left(Pe^{*}\right)^{-0.4867}+0.1763$. The experimental data agree well with the anticipated power-law relations. $F$ is plotted as $F=1/m+1$ (dashed line) where $m$ values are estimated from the power-law scaling of $m$ in (c). (d) The modified Nakagami-$m$ distribution reflects the net deposition which underscores the competition between the most probable diffusive energy and energy barrier for deposition. }
\end{figure}

We obtain the probability data (PD) for simulated $v$ by sampling the entire ensemble of particles. The long-time PD ($t=10 \tau$) for $5 \times 10^{-4} \leq Pe^{*} \leq 619$ are shown in Figs. \ref{fig:1}(a) and \ref{fig:2}(a), where $\tau=l /\langle v_{\mathrm{w}}\rangle$ yields the characteristic advection time for fluid transport through an average pore size $l$. As $Pe^{*}$ increases, the probability distributions transition from Gaussian to non-Gaussian, the latter of which features a sharp peak at near-zero velocities and heavy right tails. The numerical results agree with the experiments shown in Fig. \ref{fig:1}(b). Based on the maximum-likelihood (ML) estimation, the tail over positive $v$ is fitted with the modified Nakagami-$m$ PDF and other widely used PDFs, e.g., exponential, half-normal, log-normal, and gamma distributions \cite{RN30}. To fit the modified Nakagami-$m$ distribution, parameters $m$ and $\Omega$ are obtained by solving the derivatives of the logarithmic likelihood function \cite{RN31}. For all cases studied when $Pe^{*} \in\left[5 \times 10^{-4}, 619\right]$, out of all the functions considered, the modified Nakagami-$m$ distribution is the one most consistent with the entire range of velocity data, up to the noise floor. An example shown in Fig. \ref{fig:1}(c) at $Pe^{*}=59$ along $\mathrm{with}$ the cases at other $Pe^{*}$ values (in SM, Sec. VIII) strongly supports the universality of Eq. \eqref{eq:5}.

Based on Eq. \eqref{eq:5}, we also derive the PDF of the dimensionless total energy $\varepsilon=\chi v^{2},$ where $\chi=\rho_{\mathrm{p}}/\rho_{\mathrm{w}}$ is the particle specific gravity (see definition of $\varepsilon$ and the derived $\varepsilon$ PDF in SM, Sec. III). We use the $m$ and $\Omega$ values, estimated from the ML Nakagami-$m$ fit of $v$ data, as inputs for the $\varepsilon$ PDF. The derived $\varepsilon$ PDF is then compared against its counterpart obtained computationally for all $Pe^{*}$. A similar procedure is performed to derive other $\varepsilon$ PDFs by imposing different $v$ PDFs, and to compare their $\varepsilon$ predictions with the counterpart simulation data. Without tuning parameters, good agreement between the $\varepsilon$ PDF, derived assuming the modified Nakagami-$m$ distribution of $v$, and the simulation data. This agreement demonstrates the predictive ability of Eq. \eqref{eq:5} (more details in SM, Sec. IX).

\paragraph{Physical significance of $\Omega$ and $m$.}{\hspace{-1em}}---Analysis of the the derived function (Eq. \eqref{eq:5}) shows that the scale parameter $\Omega=\langle v^{2}\rangle=\langle v_{\mathrm{pL}}^{2}\rangle /\langle v_{\mathrm{w}}^{2}\rangle$ is related to the mean dimensionless total energy $\langle\varepsilon\rangle$, i.e., the mean total energy of particles compared to that of the ambient fluid, as $\Omega=\langle\varepsilon\rangle / \chi$. We, therefore, suggest that $\Omega$ is a \textit{transport parameter} that reflects particles' transport strength. Our simulation results show that when the particles are initially introduced into the environment, they acquire energy from the ambient fluid; consequently, $\Omega$ increases during the first characteristic advection time $(0<t \lesssim \tau)$. Thereafter, the driving force induced by the pressure gradient is compensating the drag force to reach steady states. The particles at high $Pe^{*}$ (when $v_{\mathrm{pL}}=u_{\mathrm{pL}}$ and $w_{\mathrm{pL}}=0$) quickly reach these conditions, so that $v_{\mathrm{pL}}\rightarrow\langle v_{\mathrm{w}}\rangle$ and $\Omega \rightarrow 1$. Reaching such steady-state conditions can be delayed at lower $Pe^{*}$, because diffusion dominates (when $v_{\mathrm{pL}}=w_{\mathrm{pL}}$ and $u_{\mathrm{pL}}=0$) and the correspondent disordered motion precludes $\Omega$ from reaching unity (see the simulated evolution of $\Omega$ with time in SM, Sec. VI). Therefore, consistent with this analysis, both simulated and experimental data show that $\Omega$ at low $Pe^{*}$ is higher than $\Omega$ at high $Pe^{*}$ (see Fig. \ref{fig:2}(b)).

In the original Nakagami-$m$ distribution, the shape parameter $m$ was statistically defined as $m=(\langle v^{2}\rangle / \sigma\left[v^{2}\right])^{2}$\cite{RN28} ($\sigma[\cdot]$ is standard deviation), which makes itself relevant to the flatness of the PDF ($F=\langle v^{4}\rangle/\langle v^{2}\rangle^{2}$ \cite{RN32}) via $m=1/(F-1)$. Because (1) $F$ can indicate the deviation from the Gaussian distribution \cite{RN32,RN33}, reflecting the degree of bend in a PDF curve at low velocities (e.g., $F=3$ and $m=0.5$ representing a Gaussian distribution), and (2) such velocities reflects the particles that may deposit near the wall, $F$ directly and $m$ \textit{inversely} reflect the likelihood of particle deposition, i.e., $m$ is a \textit{deposition parameter}.

Via the relation $m=\alpha/2=T/2\mathcal{P}T_{0},$ derived in Eqs. \eqref{eq:4} and \eqref{eq:5}, we find that the deposition strength ($\sim 1/m$) depends on (1) the probability ($\mathcal{P}$) that particles will reach the near-wall region and (2) the temperature ratio ($T/T_{0}$) near the pore wall. In other words:

(1) $m$ is a function of $\mathcal{P}$ in Eq. \eqref{eq:3}, i.e., pore structure (e.g., average pore size, $l$) and hydrodynamic conditions (e.g., $Pe^*$). With the decrease of $l$, the environment is more crowded, and thus more particles will reach the near-wall regions (i.e., $\mathcal{P}$ increases), promoting the particles’ deposition (i.e., $m$ decreases).

With the increase of $Pe^{*}$, particle transport is dominated by (ordered) advection rather than (disordered) diffusion; more particles will follow the fluid streamlines and reach the region near the pore wall, i.e., the probability $\mathcal{P}$ increases, therefore deposition strength increases and $m$ decreases. Via Eq. \eqref{eq:3} and $m=T/2\mathcal{P}T_{0},$ we find that theoretical $m \propto\left(Pe^{*}\right)^{-0.43}.$ Such scaling coincides with the fitted function of the simulated $m$, as $m \propto\left(Pe^{*}\right)^{-0.4867}$ in Fig. \ref{fig:2}(c), which validates our theory for particles near the pore wall. Note that both theoretical and simulated $m$ values indicate that when $Pe^{*} \geq 10$, $m$ reaches a plateau limit, indicating the deposition of the maximum strength has occurred once $Pe^{*}$ reaches an intermediate value.

(2) $m$ also depends on $T/T_0$. The temperature $T$ is related to the kinetic energy of a particle with the most probable diffusive speed ($\hat{w}_{p}$) in the MB distribution, i.e., the \textit{most probable} diffusive energy: $\hat{K}_{\mathrm{D}}=m_{\mathrm{p}} \hat{w}_{\mathrm{p}}^{2}/2=k_{\mathrm{B}} T$ (see SM, Sec. II.A). $T_{0}$ is related to the mean energy barrier, i.e., $\langle\Delta E\rangle=k_{\mathrm{B}}T_{0}$ (see discussion related to $\Delta E$ distribution). Therefore, $m=\hat{K}_{\mathrm{D}}/2\mathcal{P}\langle\Delta E\rangle,$ which indicates $m$ evaluates the competition between the most probable diffusive energy and the mean energy barrier due to the XDLVO interactions. Consider $\mathcal{P}=1$, i.e., under the circumstances that all particles were originally near the pore walls: As shown in Fig. \ref{fig:2}(d), when $m<0.5, \hat{K}_{\mathrm{D}}<\langle\Delta E\rangle$ and a net particle deposition occurs; when $m>0.5, \hat{K}_{\mathrm{D}}>\langle\Delta E\rangle$ and the frequency of particles' release is greater than that of deposition, yielding zero net deposition; when $m=0.5$, $\hat{K}_{\mathrm{D}}=\langle\Delta E\rangle$ and a kinetic equilibrium is achieved between the particles' deposition and release. In the last case, when $m=0.5$, the modified Nakagami-$m$ distribution reduces to the one-dimensional MB distribution ($\forall v\geq 0$), we, therefore, imply that the MB distribution holds not only for non-interacting particles, as traditionally believed, but also for charged particles when $\hat{K}_{\mathrm{D}}=\langle\Delta E\rangle$ and $\mathcal{P}=1$ both occur. 

Non-Gaussian distributions ($m\neq0.5$) have been traditionally attributed to fluid stagnation and are known to be pronounced in heterogeneous environments \cite{RN9,RN12,Bijeljic}. We find here that for transport with suspended particles, the impact of fluid stagnation is manifested in particle deposition. In addition, we show that fluid stagnation is not the only reason for non-Gaussian kinetic distributions; other factors that facilitate deposition, e.g., high $Pe^*$ and low $T/T_0$, may be independent of environmental structures. 

\paragraph{Conclusions.}{\hspace{-1em}}---A putative universal probability distribution function is derived to describe for particles' kinetics in disordered crowded environments. The function reveals that non-Gaussian velocity distributions are a direct consequence of particles' transport and deposition, where their transport (reflected by $\Omega$) depends on $Pe^*$ and their deposition (reflected by $m$) depends on $Pe^*$, pore structure, and the energy imbalance near pore walls. The function's universality is predicated on the basis of the theoretical description of particle's hydrodynamic interception and transport, and electrokinetic release, and is further underpinned by Lagrangian particle-tracking simulations and experimental data. The uniqueness of the function consists in its predictive ability to describe the entire ensemble including immobile through highly mobile particles. Because our analysis of particle-wall interactions is applicable to particle-particle ones, the function is generalizable to other particulate systems where different interplays of particles and environments arise, e.g., clustered granular gases that exhibit non-Gaussian kinetic distributions \cite{RN1,RN34,RN35,RN36,RN37}. 

\begin{acknowledgments}
This work is supported by the Science4CleanEnergy European research consortium funded by European Union's Horizon 2020 research and innovation programme, under Grant Agreement No. 764810 (S4CE). Generous allocations of computing time were provided by the University College London Research Computing Platforms Support (Kathleen). We are indebted to M. Holzner and M. Carrel for offering three-dimensional particle tracking velocimetry experimental data as well as V.L. Morales for helpful correspondence. 
\end{acknowledgments}

\nocite{*}

\bibliography{apssamp}

\begin{thebibliography}{49}%
\makeatletter
\providecommand \@ifxundefined [1]{%
 \@ifx{#1\undefined}
}%
\providecommand \@ifnum [1]{%
 \ifnum #1\expandafter \@firstoftwo
 \else \expandafter \@secondoftwo
 \fi
}%
\providecommand \@ifx [1]{%
 \ifx #1\expandafter \@firstoftwo
 \else \expandafter \@secondoftwo
 \fi
}%
\providecommand \natexlab [1]{#1}%
\providecommand \enquote  [1]{``#1''}%
\providecommand \bibnamefont  [1]{#1}%
\providecommand \bibfnamefont [1]{#1}%
\providecommand \citenamefont [1]{#1}%
\providecommand \href@noop [0]{\@secondoftwo}%
\providecommand \href [0]{\begingroup \@sanitize@url \@href}%
\providecommand \@href[1]{\@@startlink{#1}\@@href}%
\providecommand \@@href[1]{\endgroup#1\@@endlink}%
\providecommand \@sanitize@url [0]{\catcode `\\12\catcode `\$12\catcode
  `\&12\catcode `\#12\catcode `\^12\catcode `\_12\catcode `\%12\relax}%
\providecommand \@@startlink[1]{}%
\providecommand \@@endlink[0]{}%
\providecommand \url  [0]{\begingroup\@sanitize@url \@url }%
\providecommand \@url [1]{\endgroup\@href {#1}{\urlprefix }}%
\providecommand \urlprefix  [0]{URL }%
\providecommand \Eprint [0]{\href }%
\providecommand \doibase [0]{https://doi.org/}%
\providecommand \selectlanguage [0]{\@gobble}%
\providecommand \bibinfo  [0]{\@secondoftwo}%
\providecommand \bibfield  [0]{\@secondoftwo}%
\providecommand \translation [1]{[#1]}%
\providecommand \BibitemOpen [0]{}%
\providecommand \bibitemStop [0]{}%
\providecommand \bibitemNoStop [0]{.\EOS\space}%
\providecommand \EOS [0]{\spacefactor3000\relax}%
\providecommand \BibitemShut  [1]{\csname bibitem#1\endcsname}%
\let\auto@bib@innerbib\@empty
\bibitem [{\citenamefont {van Zon}\ and\ \citenamefont
  {MacKintosh}(2004)}]{RN1}%
  \BibitemOpen
  \bibfield  {author} {\bibinfo {author} {\bibfnamefont {J.~S.}\ \bibnamefont
  {van Zon}}\ and\ \bibinfo {author} {\bibfnamefont {F.~C.}\ \bibnamefont
  {MacKintosh}},\ }\href@noop {} {\bibfield  {journal} {\bibinfo  {journal}
  {Physical Review Letters}\ }\textbf {\bibinfo {volume} {93}},\ \bibinfo
  {pages} {038001} (\bibinfo {year} {2004})}\BibitemShut {NoStop}%
\bibitem [{\citenamefont {H\"{o}fling}\ \emph {et~al.}(2006)\citenamefont
  {H\"{o}fling}, \citenamefont {Franosch},\ and\ \citenamefont {Frey}}]{RN2}%
  \BibitemOpen
  \bibfield  {author} {\bibinfo {author} {\bibfnamefont {F.}~\bibnamefont
  {H\"{o}fling}}, \bibinfo {author} {\bibfnamefont {T.}~\bibnamefont
  {Franosch}},\ and\ \bibinfo {author} {\bibfnamefont {E.}~\bibnamefont
  {Frey}},\ }\href@noop {} {\bibfield  {journal} {\bibinfo  {journal} {Physical
  Review Letters}\ }\textbf {\bibinfo {volume} {96}},\ \bibinfo {pages}
  {165901} (\bibinfo {year} {2006})}\BibitemShut {NoStop}%
\bibitem [{\citenamefont {Carrel}\ \emph {et~al.}(2018)\citenamefont {Carrel},
  \citenamefont {Morales}, \citenamefont {Dentz}, \citenamefont {Derlon},
  \citenamefont {Morgenroth},\ and\ \citenamefont {Holzner}}]{RN3}%
  \BibitemOpen
  \bibfield  {author} {\bibinfo {author} {\bibfnamefont {M.}~\bibnamefont
  {Carrel}}, \bibinfo {author} {\bibfnamefont {V.~L.}\ \bibnamefont {Morales}},
  \bibinfo {author} {\bibfnamefont {M.}~\bibnamefont {Dentz}}, \bibinfo
  {author} {\bibfnamefont {N.}~\bibnamefont {Derlon}}, \bibinfo {author}
  {\bibfnamefont {E.}~\bibnamefont {Morgenroth}},\ and\ \bibinfo {author}
  {\bibfnamefont {M.}~\bibnamefont {Holzner}},\ }\href@noop {} {\bibfield
  {journal} {\bibinfo  {journal} {Water Resources Research}\ }\textbf {\bibinfo
  {volume} {54}},\ \bibinfo {pages} {2183} (\bibinfo {year}
  {2018})}\BibitemShut {NoStop}%
\bibitem [{\citenamefont {Kang}\ \emph {et~al.}(2014)\citenamefont {Kang},
  \citenamefont {de~Anna}, \citenamefont {Nunes}, \citenamefont {Bijeljic},
  \citenamefont {Blunt},\ and\ \citenamefont {Juanes}}]{RN4}%
  \BibitemOpen
  \bibfield  {author} {\bibinfo {author} {\bibfnamefont {P.~K.}\ \bibnamefont
  {Kang}}, \bibinfo {author} {\bibfnamefont {P.}~\bibnamefont {de~Anna}},
  \bibinfo {author} {\bibfnamefont {J.~P.}\ \bibnamefont {Nunes}}, \bibinfo
  {author} {\bibfnamefont {B.}~\bibnamefont {Bijeljic}}, \bibinfo {author}
  {\bibfnamefont {M.~J.}\ \bibnamefont {Blunt}},\ and\ \bibinfo {author}
  {\bibfnamefont {R.}~\bibnamefont {Juanes}},\ }\href@noop {} {\bibfield
  {journal} {\bibinfo  {journal} {Geophysical Research Letters}\ }\textbf
  {\bibinfo {volume} {41}},\ \bibinfo {pages} {6184} (\bibinfo {year}
  {2014})}\BibitemShut {NoStop}%
\bibitem [{\citenamefont {Tufenkji}\ and\ \citenamefont
  {Elimelech}(2004)}]{RN5}%
  \BibitemOpen
  \bibfield  {author} {\bibinfo {author} {\bibfnamefont {N.}~\bibnamefont
  {Tufenkji}}\ and\ \bibinfo {author} {\bibfnamefont {M.}~\bibnamefont
  {Elimelech}},\ }\href@noop {} {\bibfield  {journal} {\bibinfo  {journal}
  {Environmental Science \& Technology}\ }\textbf {\bibinfo {volume} {38}},\
  \bibinfo {pages} {529} (\bibinfo {year} {2004})}\BibitemShut {NoStop}%
\bibitem [{\citenamefont {Bradford}\ \emph {et~al.}(2003)\citenamefont
  {Bradford}, \citenamefont {Simunek}, \citenamefont {Bettahar}, \citenamefont
  {van Genuchten},\ and\ \citenamefont {Yates}}]{RN6}%
  \BibitemOpen
  \bibfield  {author} {\bibinfo {author} {\bibfnamefont {S.~A.}\ \bibnamefont
  {Bradford}}, \bibinfo {author} {\bibfnamefont {J.}~\bibnamefont {Simunek}},
  \bibinfo {author} {\bibfnamefont {M.}~\bibnamefont {Bettahar}}, \bibinfo
  {author} {\bibfnamefont {M.~T.}\ \bibnamefont {van Genuchten}},\ and\
  \bibinfo {author} {\bibfnamefont {S.~R.}\ \bibnamefont {Yates}},\ }\href@noop
  {} {\bibfield  {journal} {\bibinfo  {journal} {Environmental Science \&
  Technology}\ }\textbf {\bibinfo {volume} {37}},\ \bibinfo {pages} {2242}
  (\bibinfo {year} {2003})}\BibitemShut {NoStop}%
\bibitem [{\citenamefont {Tuval}\ \emph {et~al.}(2005)\citenamefont {Tuval},
  \citenamefont {Mezi\'{c}}, \citenamefont {Bottausci}, \citenamefont {Zhang},
  \citenamefont {MacDonald},\ and\ \citenamefont {Piro}}]{RN7}%
  \BibitemOpen
  \bibfield  {author} {\bibinfo {author} {\bibfnamefont {I.}~\bibnamefont
  {Tuval}}, \bibinfo {author} {\bibfnamefont {I.}~\bibnamefont {Mezi\'{c}}},
  \bibinfo {author} {\bibfnamefont {F.}~\bibnamefont {Bottausci}}, \bibinfo
  {author} {\bibfnamefont {Y.~T.}\ \bibnamefont {Zhang}}, \bibinfo {author}
  {\bibfnamefont {N.~C.}\ \bibnamefont {MacDonald}},\ and\ \bibinfo {author}
  {\bibfnamefont {O.}~\bibnamefont {Piro}},\ }\href@noop {} {\bibfield
  {journal} {\bibinfo  {journal} {Physical Review Letters}\ }\textbf {\bibinfo
  {volume} {95}},\ \bibinfo {pages} {236002} (\bibinfo {year}
  {2005})}\BibitemShut {NoStop}%
\bibitem [{\citenamefont {Datta}\ \emph {et~al.}(2013)\citenamefont {Datta},
  \citenamefont {Chiang}, \citenamefont {Ramakrishnan},\ and\ \citenamefont
  {Weitz}}]{RN8}%
  \BibitemOpen
  \bibfield  {author} {\bibinfo {author} {\bibfnamefont {S.~S.}\ \bibnamefont
  {Datta}}, \bibinfo {author} {\bibfnamefont {H.}~\bibnamefont {Chiang}},
  \bibinfo {author} {\bibfnamefont {T.~S.}\ \bibnamefont {Ramakrishnan}},\ and\
  \bibinfo {author} {\bibfnamefont {D.~A.}\ \bibnamefont {Weitz}},\ }\href@noop
  {} {\bibfield  {journal} {\bibinfo  {journal} {Physical Review Letters}\
  }\textbf {\bibinfo {volume} {111}},\ \bibinfo {pages} {064501} (\bibinfo
  {year} {2013})}\BibitemShut {NoStop}%
\bibitem [{\citenamefont {Siena}\ \emph {et~al.}(2014)\citenamefont {Siena},
  \citenamefont {Riva}, \citenamefont {Hyman}, \citenamefont {Winter},\ and\
  \citenamefont {Guadagnini}}]{RN9}%
  \BibitemOpen
  \bibfield  {author} {\bibinfo {author} {\bibfnamefont {M.}~\bibnamefont
  {Siena}}, \bibinfo {author} {\bibfnamefont {M.}~\bibnamefont {Riva}},
  \bibinfo {author} {\bibfnamefont {J.~D.}\ \bibnamefont {Hyman}}, \bibinfo
  {author} {\bibfnamefont {C.~L.}\ \bibnamefont {Winter}},\ and\ \bibinfo
  {author} {\bibfnamefont {A.}~\bibnamefont {Guadagnini}},\ }\href@noop {}
  {\bibfield  {journal} {\bibinfo  {journal} {Physical Review E}\ }\textbf
  {\bibinfo {volume} {89}},\ \bibinfo {pages} {013018} (\bibinfo {year}
  {2014})}\BibitemShut {NoStop}%
\bibitem [{\citenamefont {Matyka}\ \emph {et~al.}(2016)\citenamefont {Matyka},
  \citenamefont {Go\l{}embiewski},\ and\ \citenamefont {Koza}}]{RN10}%
  \BibitemOpen
  \bibfield  {author} {\bibinfo {author} {\bibfnamefont {M.}~\bibnamefont
  {Matyka}}, \bibinfo {author} {\bibfnamefont {J.}~\bibnamefont
  {Go\l{}embiewski}},\ and\ \bibinfo {author} {\bibfnamefont {Z.}~\bibnamefont
  {Koza}},\ }\href@noop {} {\bibfield  {journal} {\bibinfo  {journal} {Physical
  Review E}\ }\textbf {\bibinfo {volume} {93}},\ \bibinfo {pages} {013110}
  (\bibinfo {year} {2016})}\BibitemShut {NoStop}%
\bibitem [{\citenamefont {Aramideh}\ \emph {et~al.}(2018)\citenamefont
  {Aramideh}, \citenamefont {Vlachos},\ and\ \citenamefont {Ardekani}}]{RN11}%
  \BibitemOpen
  \bibfield  {author} {\bibinfo {author} {\bibfnamefont {S.}~\bibnamefont
  {Aramideh}}, \bibinfo {author} {\bibfnamefont {P.~P.}\ \bibnamefont
  {Vlachos}},\ and\ \bibinfo {author} {\bibfnamefont {A.~M.}\ \bibnamefont
  {Ardekani}},\ }\href@noop {} {\bibfield  {journal} {\bibinfo  {journal}
  {Physical Review E}\ }\textbf {\bibinfo {volume} {98}},\ \bibinfo {pages}
  {013104} (\bibinfo {year} {2018})}\BibitemShut {NoStop}%
\bibitem [{\citenamefont {de~Anna}\ \emph {et~al.}(2017)\citenamefont
  {de~Anna}, \citenamefont {Quaife}, \citenamefont {Biros},\ and\ \citenamefont
  {Juanes}}]{RN12}%
  \BibitemOpen
  \bibfield  {author} {\bibinfo {author} {\bibfnamefont {P.}~\bibnamefont
  {de~Anna}}, \bibinfo {author} {\bibfnamefont {B.}~\bibnamefont {Quaife}},
  \bibinfo {author} {\bibfnamefont {G.}~\bibnamefont {Biros}},\ and\ \bibinfo
  {author} {\bibfnamefont {R.}~\bibnamefont {Juanes}},\ }\href@noop {}
  {\bibfield  {journal} {\bibinfo  {journal} {Physical Review Fluids}\ }\textbf
  {\bibinfo {volume} {2}},\ \bibinfo {pages} {124103} (\bibinfo {year}
  {2017})}\BibitemShut {NoStop}%
\bibitem [{\citenamefont {de~Anna}\ \emph {et~al.}(2013)\citenamefont
  {de~Anna}, \citenamefont {Le~Borgne}, \citenamefont {Dentz}, \citenamefont
  {Tartakovsky}, \citenamefont {Bolster},\ and\ \citenamefont {Davy}}]{RN13}%
  \BibitemOpen
  \bibfield  {author} {\bibinfo {author} {\bibfnamefont {P.}~\bibnamefont
  {de~Anna}}, \bibinfo {author} {\bibfnamefont {T.}~\bibnamefont {Le~Borgne}},
  \bibinfo {author} {\bibfnamefont {M.}~\bibnamefont {Dentz}}, \bibinfo
  {author} {\bibfnamefont {A.~M.}\ \bibnamefont {Tartakovsky}}, \bibinfo
  {author} {\bibfnamefont {D.}~\bibnamefont {Bolster}},\ and\ \bibinfo {author}
  {\bibfnamefont {P.}~\bibnamefont {Davy}},\ }\href@noop {} {\bibfield
  {journal} {\bibinfo  {journal} {Physical Review Letters}\ }\textbf {\bibinfo
  {volume} {110}},\ \bibinfo {pages} {184502} (\bibinfo {year}
  {2013})}\BibitemShut {NoStop}%
\bibitem [{\citenamefont {Alim}\ \emph {et~al.}(2017)\citenamefont {Alim},
  \citenamefont {Parsa}, \citenamefont {Weitz},\ and\ \citenamefont
  {Brenner}}]{RN14}%
  \BibitemOpen
  \bibfield  {author} {\bibinfo {author} {\bibfnamefont {K.}~\bibnamefont
  {Alim}}, \bibinfo {author} {\bibfnamefont {S.}~\bibnamefont {Parsa}},
  \bibinfo {author} {\bibfnamefont {D.~A.}\ \bibnamefont {Weitz}},\ and\
  \bibinfo {author} {\bibfnamefont {M.~P.}\ \bibnamefont {Brenner}},\
  }\href@noop {} {\bibfield  {journal} {\bibinfo  {journal} {Physical Review
  Letters}\ }\textbf {\bibinfo {volume} {119}},\ \bibinfo {pages} {144501}
  (\bibinfo {year} {2017})}\BibitemShut {NoStop}%
\bibitem [{\citenamefont {Rouyer}\ \emph {et~al.}(1999)\citenamefont {Rouyer},
  \citenamefont {Martin},\ and\ \citenamefont {Salin}}]{RN15}%
  \BibitemOpen
  \bibfield  {author} {\bibinfo {author} {\bibfnamefont {F.}~\bibnamefont
  {Rouyer}}, \bibinfo {author} {\bibfnamefont {J.}~\bibnamefont {Martin}},\
  and\ \bibinfo {author} {\bibfnamefont {D.}~\bibnamefont {Salin}},\
  }\href@noop {} {\bibfield  {journal} {\bibinfo  {journal} {Physical Review
  Letters}\ }\textbf {\bibinfo {volume} {83}},\ \bibinfo {pages} {1058}
  (\bibinfo {year} {1999})}\BibitemShut {NoStop}%
\bibitem [{\citenamefont {Gu\'{e}don}\ \emph {et~al.}(2019)\citenamefont
  {Gu\'{e}don}, \citenamefont {Inzoli}, \citenamefont {Riva},\ and\
  \citenamefont {Guadagnini}}]{RN16}%
  \BibitemOpen
  \bibfield  {author} {\bibinfo {author} {\bibfnamefont {G.~R.}\ \bibnamefont
  {Gu\'{e}don}}, \bibinfo {author} {\bibfnamefont {F.}~\bibnamefont {Inzoli}},
  \bibinfo {author} {\bibfnamefont {M.}~\bibnamefont {Riva}},\ and\ \bibinfo
  {author} {\bibfnamefont {A.}~\bibnamefont {Guadagnini}},\ }\href@noop {}
  {\bibfield  {journal} {\bibinfo  {journal} {Physical Review E}\ }\textbf
  {\bibinfo {volume} {100}},\ \bibinfo {pages} {043101} (\bibinfo {year}
  {2019})}\BibitemShut {NoStop}%
\bibitem [{\citenamefont {Kutsovsky}\ \emph {et~al.}(1996)\citenamefont
  {Kutsovsky}, \citenamefont {Scriven}, \citenamefont {Davis},\ and\
  \citenamefont {Hammer}}]{RN17}%
  \BibitemOpen
  \bibfield  {author} {\bibinfo {author} {\bibfnamefont {Y.}~\bibnamefont
  {Kutsovsky}}, \bibinfo {author} {\bibfnamefont {L.}~\bibnamefont {Scriven}},
  \bibinfo {author} {\bibfnamefont {H.}~\bibnamefont {Davis}},\ and\ \bibinfo
  {author} {\bibfnamefont {B.~E.}\ \bibnamefont {Hammer}},\ }\href@noop {}
  {\bibfield  {journal} {\bibinfo  {journal} {Physics of Fluids}\ }\textbf
  {\bibinfo {volume} {8}},\ \bibinfo {pages} {863} (\bibinfo {year}
  {1996})}\BibitemShut {NoStop}%
\bibitem [{\citenamefont {McQuarrie}\ and\ \citenamefont {Simon}(1997)}]{RN18}%
  \BibitemOpen
  \bibfield  {author} {\bibinfo {author} {\bibfnamefont {D.~A.}\ \bibnamefont
  {McQuarrie}}\ and\ \bibinfo {author} {\bibfnamefont {J.~D.}\ \bibnamefont
  {Simon}},\ }\href@noop {} {\emph {\bibinfo {title} {Physical chemistry: A
  molecular approach}}}\ (\bibinfo  {publisher} {Sterling Publishing Company},\
  \bibinfo {year} {1997})\BibitemShut {NoStop}%
\bibitem [{\citenamefont {Kretzschmar}\ \emph {et~al.}(1999)\citenamefont
  {Kretzschmar}, \citenamefont {Borkovec}, \citenamefont {Grolimund},\ and\
  \citenamefont {Elimelech}}]{RN19}%
  \BibitemOpen
  \bibfield  {author} {\bibinfo {author} {\bibfnamefont {R.}~\bibnamefont
  {Kretzschmar}}, \bibinfo {author} {\bibfnamefont {M.}~\bibnamefont
  {Borkovec}}, \bibinfo {author} {\bibfnamefont {D.}~\bibnamefont
  {Grolimund}},\ and\ \bibinfo {author} {\bibfnamefont {M.}~\bibnamefont
  {Elimelech}},\ }\bibinfo {title} {Mobile subsurface colloids and their role
  in contaminant transport},\ in\ \href@noop {} {\emph {\bibinfo {booktitle}
  {Advances in agronomy}}},\ Vol.~\bibinfo {volume} {66}\ (\bibinfo
  {publisher} {Elsevier},\ \bibinfo {year} {1999})\ pp.\ \bibinfo {pages}
  {121--193}\BibitemShut {NoStop}%
\bibitem [{\citenamefont {Klimontovich}(1995)}]{RN20}%
  \BibitemOpen
  \bibfield  {author} {\bibinfo {author} {\bibfnamefont {Y.}~\bibnamefont
  {Klimontovich}},\ }\href {https://books.google.co.uk/books?id=WjlqCQAAQBAJ}
  {\emph {\bibinfo {title} {Statistical Theory of Open Systems: Volume 1: A
  Unified Approach to Kinetic Description of Processes in Active Systems}}},\
  Fundamental Theories of Physics\ (\bibinfo  {publisher} {Springer
  Netherlands},\ \bibinfo {year} {1995})\BibitemShut {NoStop}%
\bibitem [{\citenamefont {Pauli}\ and\ \citenamefont {Enz}(2000)}]{RN21}%
  \BibitemOpen
  \bibfield  {author} {\bibinfo {author} {\bibfnamefont {W.}~\bibnamefont
  {Pauli}}\ and\ \bibinfo {author} {\bibfnamefont {C.~P.}\ \bibnamefont
  {Enz}},\ }\href@noop {} {\emph {\bibinfo {title} {Thermodynamics and the
  kinetic theory of gases}}},\ Vol.~\bibinfo {volume} {3}\ (\bibinfo
  {publisher} {Courier Corporation},\ \bibinfo {year} {2000})\BibitemShut
  {NoStop}%
\bibitem [{\citenamefont {Civan}(2011)}]{RN22}%
  \BibitemOpen
  \bibfield  {author} {\bibinfo {author} {\bibfnamefont {F.}~\bibnamefont
  {Civan}},\ }\href {https://doi.org/10.1002/9781118086810} {\emph {\bibinfo
  {title} {Porous Media Transport Phenomena}}}\ (\bibinfo  {publisher} {John
  Wiley \& Sons},\ \bibinfo {year} {2011})\BibitemShut {NoStop}%
\bibitem [{\citenamefont {Schekochihin}(2018)}]{Schekochihin}%
  \BibitemOpen
  \bibfield  {author} {\bibinfo {author} {\bibfnamefont {A.~A.}\ \bibnamefont
  {Schekochihin}},\ }\href@noop {} {\bibinfo {title} {Lectures on kinetic
  theory of gases and statistical physics}} (\bibinfo {year}
  {2018})\BibitemShut {NoStop}%
\bibitem [{\citenamefont {Schwabl}\ and\ \citenamefont {Brewer}(2006)}]{RN23}%
  \BibitemOpen
  \bibfield  {author} {\bibinfo {author} {\bibfnamefont {F.}~\bibnamefont
  {Schwabl}}\ and\ \bibinfo {author} {\bibfnamefont {W.}~\bibnamefont
  {Brewer}},\ }\href@noop {} {\emph {\bibinfo {title} {Statistical
  Mechanics}}},\ Advanced Texts in Physics\ (\bibinfo  {publisher} {Springer
  Berlin Heidelberg},\ \bibinfo {year} {2006})\BibitemShut {NoStop}%
\bibitem [{\citenamefont {Masliyah}\ and\ \citenamefont
  {Bhattacharjee}(2006)}]{RN24}%
  \BibitemOpen
  \bibfield  {author} {\bibinfo {author} {\bibfnamefont {J.~H.}\ \bibnamefont
  {Masliyah}}\ and\ \bibinfo {author} {\bibfnamefont {S.}~\bibnamefont
  {Bhattacharjee}},\ }\href@noop {} {\emph {\bibinfo {title} {Electrokinetic
  and colloid transport phenomena}}}\ (\bibinfo  {publisher} {John Wiley \&
  Sons},\ \bibinfo {year} {2006})\BibitemShut {NoStop}%
\bibitem [{\citenamefont {Bhattacharjee}\ and\ \citenamefont
  {Datta}(2019)}]{Datta}%
  \BibitemOpen
  \bibfield  {author} {\bibinfo {author} {\bibfnamefont {T.}~\bibnamefont
  {Bhattacharjee}}\ and\ \bibinfo {author} {\bibfnamefont {S.~S.}\ \bibnamefont
  {Datta}},\ }\href@noop {} {\bibfield  {journal} {\bibinfo  {journal} {Nature
  communications}\ }\textbf {\bibinfo {volume} {10}},\ \bibinfo {pages} {1}
  (\bibinfo {year} {2019})}\BibitemShut {NoStop}%
\bibitem [{\citenamefont {Shen}\ \emph {et~al.}(2020)\citenamefont {Shen},
  \citenamefont {Jin}, \citenamefont {Zhuang}, \citenamefont {Li},\ and\
  \citenamefont {Xing}}]{RN25}%
  \BibitemOpen
  \bibfield  {author} {\bibinfo {author} {\bibfnamefont {C.}~\bibnamefont
  {Shen}}, \bibinfo {author} {\bibfnamefont {Y.}~\bibnamefont {Jin}}, \bibinfo
  {author} {\bibfnamefont {J.}~\bibnamefont {Zhuang}}, \bibinfo {author}
  {\bibfnamefont {T.}~\bibnamefont {Li}},\ and\ \bibinfo {author}
  {\bibfnamefont {B.}~\bibnamefont {Xing}},\ }\href@noop {} {\bibfield
  {journal} {\bibinfo  {journal} {Critical Reviews in Environmental Science and
  Technology}\ }\textbf {\bibinfo {volume} {50}},\ \bibinfo {pages} {244}
  (\bibinfo {year} {2020})}\BibitemShut {NoStop}%
\bibitem [{\citenamefont {Grolimund}\ and\ \citenamefont
  {Borkovec}(1999)}]{RN26}%
  \BibitemOpen
  \bibfield  {author} {\bibinfo {author} {\bibfnamefont {D.}~\bibnamefont
  {Grolimund}}\ and\ \bibinfo {author} {\bibfnamefont {M.}~\bibnamefont
  {Borkovec}},\ }\href@noop {} {\bibfield  {journal} {\bibinfo  {journal}
  {Environmental Science \& Technology}\ }\textbf {\bibinfo {volume} {33}},\
  \bibinfo {pages} {4054} (\bibinfo {year} {1999})}\BibitemShut {NoStop}%
\bibitem [{\citenamefont {Montgomery}\ and\ \citenamefont
  {Runger}(2003)}]{RN27}%
  \BibitemOpen
  \bibfield  {author} {\bibinfo {author} {\bibfnamefont {D.~C.}\ \bibnamefont
  {Montgomery}}\ and\ \bibinfo {author} {\bibfnamefont {G.~C.}\ \bibnamefont
  {Runger}},\ }\href@noop {} {\emph {\bibinfo {title} {Applied Statistics and
  Probability for Engineers}}}\ (\bibinfo  {publisher} {Wiley},\ \bibinfo
  {year} {2003})\BibitemShut {NoStop}%
\bibitem [{\citenamefont {Weber}\ and\ \citenamefont {Paddock}(1983)}]{Weber}%
  \BibitemOpen
  \bibfield  {author} {\bibinfo {author} {\bibfnamefont {M.}~\bibnamefont
  {Weber}}\ and\ \bibinfo {author} {\bibfnamefont {D.}~\bibnamefont
  {Paddock}},\ }\href@noop {} {\bibfield  {journal} {\bibinfo  {journal}
  {Journal of Colloid and Interface Science}\ }\textbf {\bibinfo {volume}
  {94}},\ \bibinfo {pages} {328} (\bibinfo {year} {1983})}\BibitemShut
  {NoStop}%
\bibitem [{\citenamefont {Nakagami}(1960)}]{RN28}%
  \BibitemOpen
  \bibfield  {author} {\bibinfo {author} {\bibfnamefont {M.}~\bibnamefont
  {Nakagami}},\ }\bibinfo {title} {The $m$-distribution—a general formula of
  intensity distribution of rapid fading},\ in\ \href@noop {} {\emph {\bibinfo
  {booktitle} {Statistical methods in radio wave propagation}}}\ (\bibinfo
  {publisher} {Elsevier},\ \bibinfo {year} {1960})\ pp.\ \bibinfo {pages}
  {3--36}\BibitemShut {NoStop}%
\bibitem [{\citenamefont {Kandhai}\ \emph {et~al.}(2002)\citenamefont
  {Kandhai}, \citenamefont {Hlushkou}, \citenamefont {Hoekstra}, \citenamefont
  {Sloot}, \citenamefont {Van~As},\ and\ \citenamefont {Tallarek}}]{RN29}%
  \BibitemOpen
  \bibfield  {author} {\bibinfo {author} {\bibfnamefont {D.}~\bibnamefont
  {Kandhai}}, \bibinfo {author} {\bibfnamefont {D.}~\bibnamefont {Hlushkou}},
  \bibinfo {author} {\bibfnamefont {A.~G.}\ \bibnamefont {Hoekstra}}, \bibinfo
  {author} {\bibfnamefont {P.~M.~A.}\ \bibnamefont {Sloot}}, \bibinfo {author}
  {\bibfnamefont {H.}~\bibnamefont {Van~As}},\ and\ \bibinfo {author}
  {\bibfnamefont {U.}~\bibnamefont {Tallarek}},\ }\href@noop {} {\bibfield
  {journal} {\bibinfo  {journal} {Physical Review Letters}\ }\textbf {\bibinfo
  {volume} {88}},\ \bibinfo {pages} {234501} (\bibinfo {year}
  {2002})}\BibitemShut {NoStop}%
\bibitem [{\citenamefont {Bijeljic}\ \emph {et~al.}(2011)\citenamefont
  {Bijeljic}, \citenamefont {Mostaghimi},\ and\ \citenamefont
  {Blunt}}]{Bijeljic}%
  \BibitemOpen
  \bibfield  {author} {\bibinfo {author} {\bibfnamefont {B.}~\bibnamefont
  {Bijeljic}}, \bibinfo {author} {\bibfnamefont {P.}~\bibnamefont
  {Mostaghimi}},\ and\ \bibinfo {author} {\bibfnamefont {M.~J.}\ \bibnamefont
  {Blunt}},\ }\href@noop {} {\bibfield  {journal} {\bibinfo  {journal}
  {Physical review letters}\ }\textbf {\bibinfo {volume} {107}},\ \bibinfo
  {pages} {204502} (\bibinfo {year} {2011})}\BibitemShut {NoStop}%
\bibitem [{\citenamefont {Seguin}\ \emph {et~al.}(1998)\citenamefont {Seguin},
  \citenamefont {Montillet},\ and\ \citenamefont {Comiti}}]{Seguin}%
  \BibitemOpen
  \bibfield  {author} {\bibinfo {author} {\bibfnamefont {D.}~\bibnamefont
  {Seguin}}, \bibinfo {author} {\bibfnamefont {A.}~\bibnamefont {Montillet}},\
  and\ \bibinfo {author} {\bibfnamefont {J.}~\bibnamefont {Comiti}},\
  }\href@noop {} {\bibfield  {journal} {\bibinfo  {journal} {Chemical
  engineering science}\ }\textbf {\bibinfo {volume} {53}},\ \bibinfo {pages}
  {3751} (\bibinfo {year} {1998})}\BibitemShut {NoStop}%
\bibitem [{\citenamefont {Skoge}\ \emph {et~al.}(2006)\citenamefont {Skoge},
  \citenamefont {Donev}, \citenamefont {Stillinger},\ and\ \citenamefont
  {Torquato}}]{Skoge}%
  \BibitemOpen
  \bibfield  {author} {\bibinfo {author} {\bibfnamefont {M.}~\bibnamefont
  {Skoge}}, \bibinfo {author} {\bibfnamefont {A.}~\bibnamefont {Donev}},
  \bibinfo {author} {\bibfnamefont {F.~H.}\ \bibnamefont {Stillinger}},\ and\
  \bibinfo {author} {\bibfnamefont {S.}~\bibnamefont {Torquato}},\ }\href@noop
  {} {\bibfield  {journal} {\bibinfo  {journal} {Physical Review E}\ }\textbf
  {\bibinfo {volume} {74}},\ \bibinfo {pages} {041127} (\bibinfo {year}
  {2006})}\BibitemShut {NoStop}%
\bibitem [{\citenamefont {Zhang}\ \emph {et~al.}(2008)\citenamefont {Zhang},
  \citenamefont {Pan}, \citenamefont {He},\ and\ \citenamefont {Pan}}]{Zhang}%
  \BibitemOpen
  \bibfield  {author} {\bibinfo {author} {\bibfnamefont {H.}~\bibnamefont
  {Zhang}}, \bibinfo {author} {\bibfnamefont {J.}~\bibnamefont {Pan}}, \bibinfo
  {author} {\bibfnamefont {X.}~\bibnamefont {He}},\ and\ \bibinfo {author}
  {\bibfnamefont {M.}~\bibnamefont {Pan}},\ }\href@noop {} {\bibfield
  {journal} {\bibinfo  {journal} {Journal of applied polymer science}\ }\textbf
  {\bibinfo {volume} {107}},\ \bibinfo {pages} {3306} (\bibinfo {year}
  {2008})}\BibitemShut {NoStop}%
\bibitem [{\citenamefont {Morales}\ \emph {et~al.}(2017)\citenamefont
  {Morales}, \citenamefont {Dentz}, \citenamefont {Willmann},\ and\
  \citenamefont {Holzner}}]{Morales}%
  \BibitemOpen
  \bibfield  {author} {\bibinfo {author} {\bibfnamefont {V.~L.}\ \bibnamefont
  {Morales}}, \bibinfo {author} {\bibfnamefont {M.}~\bibnamefont {Dentz}},
  \bibinfo {author} {\bibfnamefont {M.}~\bibnamefont {Willmann}},\ and\
  \bibinfo {author} {\bibfnamefont {M.}~\bibnamefont {Holzner}},\ }\href@noop
  {} {\bibfield  {journal} {\bibinfo  {journal} {Geophysical Research Letters}\
  }\textbf {\bibinfo {volume} {44}},\ \bibinfo {pages} {9361} (\bibinfo {year}
  {2017})}\BibitemShut {NoStop}%
\bibitem [{\citenamefont {Ma}\ \emph {et~al.}(2009)\citenamefont {Ma},
  \citenamefont {Pedel}, \citenamefont {Fife},\ and\ \citenamefont
  {Johnson}}]{Ma}%
  \BibitemOpen
  \bibfield  {author} {\bibinfo {author} {\bibfnamefont {H.}~\bibnamefont
  {Ma}}, \bibinfo {author} {\bibfnamefont {J.}~\bibnamefont {Pedel}}, \bibinfo
  {author} {\bibfnamefont {P.}~\bibnamefont {Fife}},\ and\ \bibinfo {author}
  {\bibfnamefont {W.~P.}\ \bibnamefont {Johnson}},\ }\href@noop {} {\bibfield
  {journal} {\bibinfo  {journal} {Environmental science \& technology}\
  }\textbf {\bibinfo {volume} {43}},\ \bibinfo {pages} {8573} (\bibinfo {year}
  {2009})}\BibitemShut {NoStop}%
\bibitem [{\citenamefont {Mikutis}\ \emph {et~al.}(2018)\citenamefont
  {Mikutis}, \citenamefont {Deuber}, \citenamefont {Schmid}, \citenamefont
  {Kittil\"{a}}, \citenamefont {Lobsiger}, \citenamefont {Puddu}, \citenamefont
  {Asgeirsson}, \citenamefont {Grass}, \citenamefont {Saar},\ and\
  \citenamefont {Stark}}]{Mikutis}%
  \BibitemOpen
  \bibfield  {author} {\bibinfo {author} {\bibfnamefont {G.}~\bibnamefont
  {Mikutis}}, \bibinfo {author} {\bibfnamefont {C.~A.}\ \bibnamefont {Deuber}},
  \bibinfo {author} {\bibfnamefont {L.}~\bibnamefont {Schmid}}, \bibinfo
  {author} {\bibfnamefont {A.}~\bibnamefont {Kittil\"{a}}}, \bibinfo {author}
  {\bibfnamefont {N.}~\bibnamefont {Lobsiger}}, \bibinfo {author}
  {\bibfnamefont {M.}~\bibnamefont {Puddu}}, \bibinfo {author} {\bibfnamefont
  {D.~O.}\ \bibnamefont {Asgeirsson}}, \bibinfo {author} {\bibfnamefont
  {R.~N.}\ \bibnamefont {Grass}}, \bibinfo {author} {\bibfnamefont {M.~O.}\
  \bibnamefont {Saar}},\ and\ \bibinfo {author} {\bibfnamefont {W.~J.}\
  \bibnamefont {Stark}},\ }\href@noop {} {\bibfield  {journal} {\bibinfo
  {journal} {Environmental science \& technology}\ }\textbf {\bibinfo {volume}
  {52}},\ \bibinfo {pages} {12142} (\bibinfo {year} {2018})}\BibitemShut
  {NoStop}%
\bibitem [{\citenamefont {Happel}\ and\ \citenamefont
  {Brenner}(2012)}]{Happel}%
  \BibitemOpen
  \bibfield  {author} {\bibinfo {author} {\bibfnamefont {J.}~\bibnamefont
  {Happel}}\ and\ \bibinfo {author} {\bibfnamefont {H.}~\bibnamefont
  {Brenner}},\ }\href@noop {} {\emph {\bibinfo {title} {Low Reynolds number
  hydrodynamics: with special applications to particulate media}}},\
  Vol.~\bibinfo {volume} {1}\ (\bibinfo  {publisher} {Springer Science \&
  Business Media},\ \bibinfo {year} {2012})\BibitemShut {NoStop}%
\bibitem [{\citenamefont {Holzner}\ \emph {et~al.}(2015)\citenamefont
  {Holzner}, \citenamefont {Morales}, \citenamefont {Willmann},\ and\
  \citenamefont {Dentz}}]{Holzner}%
  \BibitemOpen
  \bibfield  {author} {\bibinfo {author} {\bibfnamefont {M.}~\bibnamefont
  {Holzner}}, \bibinfo {author} {\bibfnamefont {V.~L.}\ \bibnamefont
  {Morales}}, \bibinfo {author} {\bibfnamefont {M.}~\bibnamefont {Willmann}},\
  and\ \bibinfo {author} {\bibfnamefont {M.}~\bibnamefont {Dentz}},\
  }\href@noop {} {\bibfield  {journal} {\bibinfo  {journal} {Physical Review
  E}\ }\textbf {\bibinfo {volume} {92}},\ \bibinfo {pages} {013015} (\bibinfo
  {year} {2015})}\BibitemShut {NoStop}%
\bibitem [{\citenamefont {James}(2006)}]{RN30}%
  \BibitemOpen
  \bibfield  {author} {\bibinfo {author} {\bibfnamefont {F.}~\bibnamefont
  {James}},\ }\href@noop {} {\emph {\bibinfo {title} {Statistical methods in
  experimental physics}}}\ (\bibinfo  {publisher} {World Scientific Publishing
  Company},\ \bibinfo {year} {2006})\BibitemShut {NoStop}%
\bibitem [{\citenamefont {Cheng}\ and\ \citenamefont {Beaulieu}(2001)}]{RN31}%
  \BibitemOpen
  \bibfield  {author} {\bibinfo {author} {\bibfnamefont {J.}~\bibnamefont
  {Cheng}}\ and\ \bibinfo {author} {\bibfnamefont {N.~C.}\ \bibnamefont
  {Beaulieu}},\ }\href@noop {} {\bibfield  {journal} {\bibinfo  {journal} {IEEE
  Communications Letters}\ }\textbf {\bibinfo {volume} {5}},\ \bibinfo {pages}
  {101} (\bibinfo {year} {2001})}\BibitemShut {NoStop}%
\bibitem [{\citenamefont {Olafsen}\ and\ \citenamefont {Urbach}(1999)}]{RN32}%
  \BibitemOpen
  \bibfield  {author} {\bibinfo {author} {\bibfnamefont {J.~S.}\ \bibnamefont
  {Olafsen}}\ and\ \bibinfo {author} {\bibfnamefont {J.~S.}\ \bibnamefont
  {Urbach}},\ }\href@noop {} {\bibfield  {journal} {\bibinfo  {journal}
  {Physical Review E}\ }\textbf {\bibinfo {volume} {60}},\ \bibinfo {pages}
  {R2468} (\bibinfo {year} {1999})}\BibitemShut {NoStop}%
\bibitem [{\citenamefont {Baxter}\ and\ \citenamefont {Olafsen}(2003)}]{RN33}%
  \BibitemOpen
  \bibfield  {author} {\bibinfo {author} {\bibfnamefont {G.}~\bibnamefont
  {Baxter}}\ and\ \bibinfo {author} {\bibfnamefont {J.~S.}\ \bibnamefont
  {Olafsen}},\ }\href@noop {} {\bibfield  {journal} {\bibinfo  {journal}
  {Nature}\ }\textbf {\bibinfo {volume} {425}},\ \bibinfo {pages} {680}
  (\bibinfo {year} {2003})}\BibitemShut {NoStop}%
\bibitem [{\citenamefont {Harth}\ \emph {et~al.}(2018)\citenamefont {Harth},
  \citenamefont {Trittel}, \citenamefont {Wegner},\ and\ \citenamefont
  {Stannarius}}]{RN34}%
  \BibitemOpen
  \bibfield  {author} {\bibinfo {author} {\bibfnamefont {K.}~\bibnamefont
  {Harth}}, \bibinfo {author} {\bibfnamefont {T.}~\bibnamefont {Trittel}},
  \bibinfo {author} {\bibfnamefont {S.}~\bibnamefont {Wegner}},\ and\ \bibinfo
  {author} {\bibfnamefont {R.}~\bibnamefont {Stannarius}},\ }\href@noop {}
  {\bibfield  {journal} {\bibinfo  {journal} {Physical Review Letters}\
  }\textbf {\bibinfo {volume} {120}},\ \bibinfo {pages} {214301} (\bibinfo
  {year} {2018})}\BibitemShut {NoStop}%
\bibitem [{\citenamefont {Prevost}\ \emph {et~al.}(2002)\citenamefont
  {Prevost}, \citenamefont {Egolf},\ and\ \citenamefont {Urbach}}]{RN35}%
  \BibitemOpen
  \bibfield  {author} {\bibinfo {author} {\bibfnamefont {A.}~\bibnamefont
  {Prevost}}, \bibinfo {author} {\bibfnamefont {D.~A.}\ \bibnamefont {Egolf}},\
  and\ \bibinfo {author} {\bibfnamefont {J.~S.}\ \bibnamefont {Urbach}},\
  }\href@noop {} {\bibfield  {journal} {\bibinfo  {journal} {Physical Review
  Letters}\ }\textbf {\bibinfo {volume} {89}},\ \bibinfo {pages} {084301}
  (\bibinfo {year} {2002})}\BibitemShut {NoStop}%
\bibitem [{\citenamefont {Olafsen}\ and\ \citenamefont {Urbach}(1998)}]{RN36}%
  \BibitemOpen
  \bibfield  {author} {\bibinfo {author} {\bibfnamefont {J.~S.}\ \bibnamefont
  {Olafsen}}\ and\ \bibinfo {author} {\bibfnamefont {J.~S.}\ \bibnamefont
  {Urbach}},\ }\href@noop {} {\bibfield  {journal} {\bibinfo  {journal}
  {Physical Review Letters}\ }\textbf {\bibinfo {volume} {81}},\ \bibinfo
  {pages} {4369} (\bibinfo {year} {1998})}\BibitemShut {NoStop}%
\bibitem [{\citenamefont {Yu}\ \emph {et~al.}(2020)\citenamefont {Yu},
  \citenamefont {Schr\"{o}ter},\ and\ \citenamefont {Sperl}}]{RN37}%
  \BibitemOpen
  \bibfield  {author} {\bibinfo {author} {\bibfnamefont {P.}~\bibnamefont
  {Yu}}, \bibinfo {author} {\bibfnamefont {M.}~\bibnamefont {Schr\"{o}ter}},\
  and\ \bibinfo {author} {\bibfnamefont {M.}~\bibnamefont {Sperl}},\
  }\href@noop {} {\bibfield  {journal} {\bibinfo  {journal} {Physical Review
  Letters}\ }\textbf {\bibinfo {volume} {124}},\ \bibinfo {pages} {208007}
  (\bibinfo {year} {2020})}\BibitemShut {NoStop}%
\end{thebibliography}%

\end{document}